\documentclass[aps,pra,
reprint,superscriptaddress,longbibliography,floatfix,showkeys]{revtex4-2}
\usepackage{amsmath,amssymb,amsthm,graphicx,bm}
\usepackage[colorlinks=true,citecolor=blue,linkcolor=blue,urlcolor=blue]{hyperref}
\usepackage{booktabs}
\newtheorem{theorem}{Theorem}
\newtheorem{proposition}{Proposition}
\newtheorem{corollary}{Corollary}

\newcommand{\Tr}{\operatorname{Tr}}
\newcommand{\Prob}{\mathbb{P}}
\newcommand{\Ex}{\mathbb{E}}

\newcommand{\id}{\mathbb{I}}
\newcommand{\e}{\mathrm{e}}
\begin{document}
\title{Decoherent Histories from Local Spectral Dephasing in Quantum Many-Body Systems}
\author{Jie Gu}
\email{jiegu1989@gmail.com}
\affiliation{Chengdu Academy of Educational Sciences, Chengdu 610036, China}
\begin{abstract}
We study local spectral dephasing as a mechanism for decoherent histories in isolated quantum many-body systems. We prove that vanishing projector-weighted spectral concentration is asymptotically equivalent to vanishing time-averaged local memory. With independently sampled waiting times and suitable conditions on the initial state and history weights, this local property suppresses normalized interference across complete fixed-length history families with high probability. For translation-invariant lattices, spectral estimates provide conditions under which dephasing occurs at long times. Exact diagonalization and sparse time evolution of a spin chain illustrate the results.
\end{abstract}
\maketitle

%\tableofcontents

\section{Introduction}

An isolated quantum many-body system evolves unitarily, yet a classical description assigns probabilities to sequences of events. The decoherent (or consistent) histories framework makes this coexistence precise without treating an external observer or wave-function collapse as fundamental ingredients \cite{Griffiths1984,Omnes1992,GellMannHartle1993,Halliwell1995Review,Hohenberg2010}. At successive times, one specifies exhaustive, mutually exclusive alternatives. Each resulting history has a chain operator, and the decoherence functional measures interference between histories \cite{DowkerHalliwell1992,Griffiths1996}. Its diagonal entries give candidate probabilities. When off-diagonal interference is negligible, these probabilities obey the usual sum rules under coarse graining. The condition concerns entire sequences of events. Relaxation of an observable at one time does not ensure consistent probabilities for a multitime history family, and such probabilities alone do not imply classical predictability or a memoryless stochastic process \cite{GellMannHartle1993,Strasberg2023,Deutsch1991,Rigol2008,GogolinEisert2016}. We therefore ask how a unitary many-body evolution can suppress interference between alternative histories.

Several mechanisms can suppress interference between histories. Environmental coupling entangles alternatives with unobserved degrees of freedom and can produce environment-induced decoherence \cite{JoosZeh1985,PazZurek1993,Zurek1981,Zurek2003,Schlosshauer2005}. Information about the alternatives may also persist in stable or redundantly accessible records \cite{Halliwell1999,OllivierPoulinZurek2004,Zurek2009,GellMannHartle2014}. In a closed many-body system, degrees of freedom excluded from the chosen histories can have a similar effect. A different route uses coarse-grained, approximately conserved densities, whose slow evolution supports hydrodynamic histories \cite{BrunHalliwell1996,Halliwell1998,Halliwell1999Local,Halliwell2003,HohenbergHalperin1977}. Direct calculations in random-matrix models and isolated spin chains have found suppressed history interference for selected observables in chaotic and interacting integrable dynamics \cite{StrasbergReinhardSchindler2024,WangStrasberg2025}. Recording and monitored models clarify how access to information about past events relates to decoherence \cite{ArefyevaPolyakov2026,FerteFarciCao2026}. Multitime equilibration poses a related question about process statistics \cite{Dowling2023Quantum,Dowling2023SciPost}. These results suggest examining the intrinsic dynamics of the local observables used to define histories. Such observables contain components at different Bohr frequencies, and their relative phases evolve at different rates. Can the resulting dephasing suppress interference between multitime histories?

We investigate local spectral dephasing as a mechanism for decoherent histories. A local projector probes a matrix-element-weighted distribution of Bohr frequencies. Dispersion of their phases can weaken its memory of earlier values. Our first theorem shows that, along sequences of systems and observation scales, vanishing weighted spectral concentration in windows of width $1/T$ is equivalent to vanishing absolute time-averaged connected autocorrelation. Spectral nonconcentration and local memory loss are the frequency- and time-domain criteria for the same asymptotic property. Our second theorem shows that weak local memory suffices to suppress normalized interference between all distinct histories in a predetermined family, in probability, when waiting times are sampled independently and the initial state and reference branch weights satisfy the stated conditions. We also establish a static-lattice realization. For fixed-support projectors in translation-invariant systems, nondegenerate energies and sparse nonzero-gap multiplicities guarantee local memory loss along suitably long observation windows. Together these results connect a local spectral property to a consistent multitime probability description.

Section~\ref{sec:mechanism} develops the physical mechanism, proves the spectral equivalence, and establishes the sufficient condition for suppression of interference across a complete history family. Section~\ref{sec:static} derives a static-lattice realization and illustrates it numerically in a one-dimensional spin chain. Section~\ref{sec:literature} compares this mechanism with other approaches to history decoherence, and Sec.~\ref{sec:conclusion} summarizes the results and open questions. The main text gives proof sketches, with full proofs in Appendices~\ref{app:proof-spectral}--\ref{app:proof-static}. Appendix~\ref{sec:dual} treats dual-unitary Floquet circuits.

\section{Local spectral dephasing as a mechanism for decoherent histories}
\label{sec:mechanism}
\subsection{Physical picture}

We give an overview of the mechanism.
The mechanism considered here is {local spectral
dephasing}. By this term, we mean the suppression of
local temporal correlations through phase cancellation
among the different frequency components of a local
observable. More specifically, after subtracting the
stationary mean of a local projector, its time-dependent
autocorrelation can be expressed as a weighted sum of
oscillations at the Bohr frequencies of the system.
During unitary evolution, these components accumulate
different phases, and their contributions can increasingly
cancel. Local spectral dephasing occurs when this
cancellation makes the magnitude of the local correlation
small on average over the observation window.
It therefore describes a loss of local dynamical memory,
rather than a loss of quantum coherence in the full
isolated system.

An important feature of this mechanism is that its
effectiveness depends on the distribution of spectral
weight, rather than simply on the number of distinct
frequencies. Components with nearly equal frequencies
accumulate similar phases and may continue to contribute
coherently over a finite observation time.
By contrast, sufficiently dispersed frequencies allow
more effective phase cancellation.
The relevant criterion is therefore whether appreciable
spectral weight remains concentrated within frequency
intervals of width set by the inverse observation time.
This finite-time viewpoint also allows for occasional
recurrences: effective dephasing requires weak
time-averaged correlations, not monotonic decay at
every instant.

To understand how this local mechanism affects history
interference, consider two histories that differ at some
observation but share all subsequent outcomes.
At their last differing observation, the corresponding
projectors are orthogonal, and the overlap between the
two history prefixes vanishes.
Unitary evolution alone preserves this orthogonality.
However, a subsequent projection onto a common outcome
can produce a nonzero overlap between the two branches,
allowing interference to reappear.
The identity component of the later projector cannot
generate such an overlap, since it only multiplies
the overlap already present.
Any regenerated interference must therefore arise
from the remaining, traceless part of that projector.
Local spectral dephasing suppresses the contribution
of this part for typical waiting times, thereby
limiting the interference that subsequent projections
can generate.

For a complete family of histories, the argument must
also account for correlations established by preceding
observations. Different earlier outcomes generally lead
to different operators entering the next stage of
evolution, so successive observations cannot be treated
as statistically independent.
Nevertheless, once a particular history prefix is fixed,
an independently sampled waiting time allows the same
local two-time estimate to control the next projection.
The identity component of the projector propagates
a reference branch weight determined by its normalized
rank, while the traceless component produces deviations
from that weight and can contribute to interference
between branches.
Under suitable conditions on the initial state, these
contributions can be controlled uniformly over the
history family.
Positive reference branch weights are additionally
required to convert bounds on the overlaps into bounds
on normalized interference.

The resulting mechanism has two distinct steps.
First, we establish that spectral nonconcentration
on the frequency scale set by the observation window
is asymptotically equivalent to weak time-averaged
local memory.
Second, under independent waiting-time sampling and
the stated conditions on the initial state and branch
weights, weak local memory provides a sufficient
condition for suppressing interference throughout a
prescribed fixed-length history family with high
probability.
The following subsections establish these two results
quantitatively.

\subsection{A finite-resolution criterion for spectral dephasing}\label{sec:spectral}
We work in a finite-dimensional Hilbert space $\mathcal H$ of dimension $D$, with a time-independent Hamiltonian $H$ and evolution $U(t)=\exp(-\mathrm{i}Ht)$, in units where $\hbar=1$. A local projector acts nontrivially only on a bounded set of sites and as the identity elsewhere.

For a chosen local projector, we characterize dephasing through its absolute time-averaged connected autocorrelation and the concentration of its spectral weight within frequency windows resolved by the observation time.

For any projector $P$ and observation-window length $T>0$, write its normalized rank as $q_P=D^{-1}\Tr P$, and define its centered form by subtracting the infinite-temperature mean, $A_P=P-q_P\id$.
Its Heisenberg evolution is $A_P(t)=U(t)^\dagger A_PU(t)$.
The connected autocorrelation $c_P(t)$, the absolute window-memory parameter $h_P(T)$, and the infinite-temperature variance $v_P$ are

\begin{equation}\label{eq:mixing}
\begin{aligned}
c_P(t)&=\frac{1}{D}\Tr[A_P(t)A_P],\\
h_P(T)&=\frac{1}{T}\int_{-T}^{T}|c_P(t)|\,\mathrm{d}t,\\
v_P&=q_P(1-q_P),
\end{aligned}
\end{equation}
For Hermitian $A_P$, stationarity gives $c_P(-t)=c_P(t)^*$. Cyclicity of the trace also makes $c_P(t)$ real and even.
At zero delay, $c_P(0)=D^{-1}\Tr A_P^2=q_P(1-q_P)=v_P\le1/4$.
The prefactor $1/T$ makes $h_P(T)$ twice the conventional average over $[-T,T]$. This normalization simplifies the time-frame estimate below and leaves the vanishing criterion unchanged.

The absolute value in $h_P(T)$ excludes cancellation between positive and negative oscillations.
Small $h_P(T)$ means that the magnitude of the local two-time correlation is small for most delays in the observation window.
Correlations may recur at particular delays, and other observables or higher-order operations may retain memory.
When considering a sequence of systems and waiting-time scales, we speak of local dynamical memory loss when the maximum of $h_P(T)$ over the projectors under consideration tends to zero.

To determine when this memory is small, we examine which Bohr frequencies carry appreciable matrix-element weight.
The matrix-element weights matter because energy gaps weakly coupled to the chosen projector contribute little to its dynamics.

Choose an orthonormal energy eigenbasis with $H|m\rangle=E_m|m\rangle$.
Write $A_{ml}=\langle m|A_P|l\rangle$, and let $\delta_\omega$ denote the unit point mass at frequency $\omega$.
We define a positive measure on frequency space and express the correlation as its Fourier transform:

\begin{equation}\label{eq:mu}
\begin{aligned}
\mu_P&=\frac1D\sum_{m,l}|\langle m|A_P|l\rangle|^2\delta_{E_m-E_l},\\
c_P(t)&=\int_{\mathbb R}\e^{\mathrm{i}\omega t}\,\mu_P(\mathrm{d}\omega).
\end{aligned}
\end{equation}
The measure weights the ordered transition $(m,l)$ by $D^{-1}|A_{ml}|^2$ and places it at its Bohr frequency $E_m-E_l$.
Repeated gaps therefore contribute to the same atom, with their weights added.
Hermiticity makes the measure symmetric under $\omega\mapsto-\omega$. Its total mass is $\mu_P(\mathbb R)=D^{-1}\Tr A_P^2=v_P$.
These facts also explain why the autocorrelation in Eq.~\eqref{eq:mixing} is real and even.

A finite observation window cannot sharply distinguish nearby frequencies.
For $T>0$, we quantify the largest locally weighted spectral mass in any closed frequency interval of width $1/T$ by

\begin{equation}\label{eq:M}
M_P(T)=\sup_{\Omega\in\mathbb R}\mu_P\bigl([\Omega,\Omega+T^{-1}]\bigr).
\end{equation}
The supremum ranges over all locations of frequency windows, including windows near zero and around any nonzero Bohr frequency.
Thus $M_P(T)$ is a concentration measure: it is large when significant projector-weighted transitions remain unresolved within one frequency window and small when no such window carries much mass.
As $T$ increases, the window narrows and $M_P(T)$ is nonincreasing, although the memory average $h_P(T)$ need not be monotone.
This weighted finite-resolution measure probes a different spectral feature from the level statistics used to characterize many-body chaos \cite{Srednicki1994,DAlessio2016}.
The next theorem shows that it is exactly the spectral question relevant to the absolute memory parameter.

\begin{theorem}[Finite-resolution spectral-dephasing criterion]\label{thm:spectrum}
For every finite-dimensional Hamiltonian, projector $P$, and observation scale $T>0$, define the numerical constant $a_0=\left[\sin(1/4)/(1/4)\right]^2$.
The window concentration $M_P(T)$ and the absolute time-averaged memory $h_P(T)$ obey the two-sided estimate

\begin{equation}\label{eq:specbound}
a_0 M_P(T)\le h_P(T)\le\sqrt{8(1+\pi^2)v_P M_P(T)}.
\end{equation}
Consequently, for a sequence of systems of dimensions $D$, with projectors $P_D$ and observation scales $T_D$, one has $h_{P_D}(T_D)\to0$ if and only if $M_{P_D}(T_D)\to0$.
The equivalence concerns joint sequences of systems and observation scales, with finite systems treated at each step.

\end{theorem}

\begin{proof}[Proof sketch]
Over a duration $T$, nearby Bohr frequencies accumulate almost the same relative phase, while well-separated ones can dephase.
A positive triangular time filter detects any unresolved spectral cluster of width $1/T$, forcing a lower bound on the average magnitude of the autocorrelation.
Conversely, applying the same filter to the squared correlation suppresses contributions from well-separated frequencies, and the remaining nearby weight is controlled by $M_P(T)$.
Cauchy--Schwarz converts this into the upper bound.
The argument permits degenerate gaps. Appendix~\ref{app:proof-spectral} gives the constants and full proof.
\end{proof}

Theorem~\ref{thm:spectrum} gives two criteria for local spectral dephasing. Spectral nonconcentration, measured by $M_P(T)$, describes the frequency distribution. Local dynamical memory loss, measured by $h_P(T)$, describes the corresponding time-averaged correlation. They have the same vanishing limit along the sequences in the theorem.
The underlying physical process is cancellation among Bohr-frequency components. At finite $T$, unresolved clusters can retain memory, and the two measures generally take different values.
This is a finite-resolution version of the spectral viewpoint behind the Wiener criterion for time-averaged mixing \cite{Katznelson2004}.

The order of limits matters.
At any fixed finite $D$, the measure is pure point and, as $T\to\infty$, $M_P(T)$ approaches the mass of its largest atom.
A vanishing sequence of window maxima can instead arise as system size increases, with the observation window also chosen as a function of size.
A Hilbert space of dimension $D$ has at most $D^2$ ordered energy-level transitions. Dephasing depends on the weighted mass in each resolved frequency window, which a bare count of distinct transitions does not capture.
In particular, the zero-frequency atom represents the component that never dephases, while a near-resonant group of nonzero frequencies can matter at experimentally finite resolution.

%Figure~\ref{fig:spectral} compares spectral concentration and local memory for the periodic chain defined in Sec.~\ref{sec:static}. The equivalence in Theorem~\ref{thm:spectrum} applies beyond this example.

\subsection{From spectral dephasing to whole-family interference suppression}\label{sec:main}
The spectral criterion controls the two-time memory of an individual projector. To apply it to multitime histories, we specify a predetermined family of projective alternatives and quantify interference between its branches.

Consider $n$ observation times $t_0=0<t_1<\cdots<t_{n-1}$, separated by waiting times $\tau_k=t_k-t_{k-1}$ for $1\le k\le n-1$.
At each observation $k$, outcomes in a finite set $\mathcal A_k$ correspond to a complete orthogonal projective decomposition $\mathcal P^{(k)}=\{P_a^{(k)}\}_{a\in\mathcal A_k}$, with $P_a^{(k)}P_b^{(k)}=0$ for $a\ne b$ and $\sum_{a\in\mathcal A_k}P_a^{(k)}=\id$.
These decompositions are fixed before the waiting times are sampled. They can describe local measurements whose projectors need not be energy eigenprojectors.
A history $\alpha=(a_0,\ldots,a_{n-1})$ selects one outcome at each observation. Its chain operator and its decoherence-functional overlap with a second history $\beta$ are
\begin{equation}\label{eq:history}
\begin{aligned}
C_\alpha&=P^{(n-1)}_{a_{n-1}}U(\tau_{n-1})\cdots U(\tau_1)P^{(0)}_{a_0},\\
D_{\alpha\beta}&=\Tr(C_\alpha\rho C_\beta^\dagger),
\end{aligned}
\end{equation}
where $\rho$ is the initial density matrix.
The matrix $(D_{\alpha\beta})$ is positive semidefinite, since the operators $C_\alpha\rho^{1/2}$ give a Gram representation.
Thus $p_\alpha=D_{\alpha\alpha}\ge0$ is the weight of a fine-grained history and $|D_{\alpha\beta}|^2\le p_\alpha p_\beta$.
For $\alpha\ne\beta$, define the normalized interference $\epsilon_{\alpha\beta}=D_{\alpha\beta}/\sqrt{p_\alpha p_\beta}$ when $p_\alpha p_\beta>0$, and set it to zero otherwise. A zero-weight history has zero overlap with every other history.
Our approximate medium-decoherence criterion is $|\epsilon_{\alpha\beta}|\le\delta$ for every distinct pair.
Normalization matters: a small unnormalized overlap need not imply weak interference relative to the corresponding branch probabilities.

We now use the local-memory bound to control every pair of histories, including after conditioning on arbitrary earlier outcomes.

For this multitime statement, we specify the probabilistic setting as follows.
Each waiting time $\tau_k$, $1\le k\le n-1$, is sampled independently and uniformly from $[0,T]$. The Hamiltonian and projectors remain fixed.
The initial state satisfies $\rho\le(\kappa/D)\id$ for some $\kappa\ge1$, equivalently $D\|\rho\|_\infty\le\kappa$, where $\|\cdot\|_\infty$ denotes the operator norm.
Let $r_{a_0}=\Tr(\rho P_{a_0}^{(0)})$ be the probability of the first outcome, and let $q_a^{(k)}=D^{-1}\Tr P_a^{(k)}$ be the normalized rank of a later outcome projector.
We introduce the reference product weights and their minimum:

\begin{equation}\label{eq:Q}
Q_\alpha=r_{a_0}\prod_{k=1}^{n-1}q_{a_k}^{(k)},\qquad Q_*=\min_\alpha Q_\alpha>0.
\end{equation}
The $Q_\alpha$ serve as comparison weights for the actual probabilities $p_\alpha$.
They describe the distribution that would result if, after the first observation, each later outcome occurred with probability given by the normalized projector rank.
Because the decompositions are complete, $\sum_\alpha Q_\alpha=1$.
Our whole-family theorem assumes $Q_*>0$, so every branch to be controlled has a nonzero reference weight.
Very small positive reference weights enter the final estimate through $Q_*^{-2}$.
The theorem below concerns complete history families for which every reference branch weight is strictly positive, i.e., $Q_*>0$.

The condition on $\rho$ controls its largest eigenvalue, which is stronger than an entropy condition.
For example, $\rho=\id/D$ allows $\kappa=1$, whereas a pure initial state requires $\kappa\ge D$ and generally makes the resulting estimate ineffective.
These assumptions have separate roles: the state bound controls the Hilbert--Schmidt norm of every history prefix, while an independent new waiting time allows the same two-time estimate to be applied after conditioning on that prefix.
The lower bound on $Q_*$ is needed at the final step, when unnormalized overlaps are converted to normalized interference.

Let $N_n=\prod_{k=0}^{n-1}|\mathcal A_k|$ be the number of fine-grained histories and $K_n=N_n(N_n-1)$ the number of ordered pairs of distinct histories.
To obtain a single bound that applies to every pair, we must control all projectors that can occur at any positive observation time.
The required uniform local-memory parameter is

\begin{equation}\label{eq:hm}
h(T)=\max_{1\le k\le n-1}\max_{a\in\mathcal A_k} h_{P_a^{(k)}}(T).
\end{equation}

\begin{theorem}[Local-memory bound for the complete history family]\label{thm:whole}
For the predetermined projective decompositions and independent uniform waiting times described above, let $n\ge2$, $Q_*>0$, and $\rho\le\kappa\id/D$.
For any $\delta>0$, the probability that at least one pair of distinct histories has normalized interference greater than $\delta$ satisfies

\begin{equation}\label{eq:mainbound}
\Prob\left(\max_{\alpha\ne\beta}|\epsilon_{\alpha\beta}|>\delta\right)
\le\frac{4\kappa(n-1)^2h(T)}{Q_*^2}\left(N_n+\frac{K_n}{\delta^2}\right).
\end{equation}
Outside an exceptional event with probability bounded by the right side of Eq.~\eqref{eq:mainbound}, every history satisfies $p_\alpha\ge Q_*/2$ and every unequal pair satisfies $|\epsilon_{\alpha\beta}|\le\delta$ simultaneously.
The probability is taken over the sampled waiting times.

\end{theorem}

\begin{proof}[Proof sketch]
Split each later projector into its scalar part, $q_a^{(k)}\id$, and its centered, phase-sensitive part $A_{P_a^{(k)}}$.
The scalar part propagates the reference branch weights. A Hilbert--Schmidt time-frame estimate bounds the conditional mean square of the centered contribution by $\kappa h(T)$ for every earlier history prefix.
When two paths choose different outcomes, their prefix trace is exactly zero. After the last disagreement, any regenerated overlap comes from small centered contributions weighted by factors $q_a^{(k)}\le1$.
At most $n-1$ such contributions control both the branch-weight errors and the off-diagonal overlaps.
A union bound then gives simultaneous control after normalizing by the positive branch weights.
Appendix~\ref{app:proof-whole} gives the operator estimate, recursion, and probability calculation.
\end{proof}

The estimate concerns a fixed Hamiltonian and measurement family with independent random waiting times. Its right side can exceed one at finite size. For a sequence of systems, the full right side must tend to zero to yield a high-probability conclusion. At fixed $n$, uniformly bounded $\kappa$, a positive lower bound on $Q_*$, and $h(T_D)\to0$ suffice. Indeed, $\sum_\alpha Q_\alpha=1$ then bounds $N_n$ and $K_n$. For growing history length, the factors $N_n$, $K_n$, $Q_*$, and $\kappa$ must be tracked explicitly. A right side below one also ensures the existence of at least one deterministic schedule satisfying the conclusions.

The same event controls probability additivity under coarse graining.
For a subset $S$ of the fine-grained histories, let $p_{\rm coh}(S)=\sum_{\alpha,\beta\in S}D_{\alpha\beta}$ be the weight of the coherent sum of their chain operators.
The sum of the individual history weights is $\sum_{\alpha\in S}p_\alpha$.
If the normalized overlap bound holds for every unequal pair, their difference obeys

\begin{equation}\label{eq:additivity}
\left|p_{\rm coh}(S)-\sum_{\alpha\in S}p_\alpha\right|\le\delta(|S|-1)\sum_{\alpha\in S}p_\alpha.
\end{equation}
Indeed, expanding $p_{\rm coh}(S)$ leaves only the off-diagonal terms in the difference.
Each satisfies $|D_{\alpha\beta}|\le\delta\sqrt{p_\alpha p_\beta}$, and $2\sqrt{xy}\le x+y$ implies $\sum_{\alpha\ne\beta\in S}\sqrt{p_\alpha p_\beta}\le(|S|-1)\sum_{\alpha\in S}p_\alpha$.
Equation~\eqref{eq:additivity} holds for every subset $S$ without requiring comparable branch weights.
Thus the same event provides approximate additivity under arbitrary coarse graining formed by grouping branches of the chosen family.

For perspective, when $\rho=\id/D$ and there are only two observation times ($n=2$), any distinct histories differ at the first or final observation and their overlap vanishes exactly by projector orthogonality.
The first nontrivial interference tests in this case have at least three observation times, so that two distinct histories can share their endpoints while differing at an intermediate observation.
This is precisely the type of interference to which the whole-family theorem applies.

Combining Theorem~\ref{thm:spectrum} with the multitime bound in Theorem~\ref{thm:whole} now gives a direct spectral-to-history criterion.
For this purpose, define the largest spectral concentration and variance over all projectors used at times $k\ge1$:
\begin{equation}\label{eq:localM}
\begin{aligned}
\mathcal M(T)&=\max_{1\le k\le n-1}\max_{a\in\mathcal A_k}M_{P_a^{(k)}}(T),\\
v_{\mathrm{loc}}&=\max_{1\le k\le n-1}\max_{a\in\mathcal A_k}v_{P_a^{(k)}}\le\tfrac14.
\end{aligned}
\end{equation}
For every positive-time projector, Theorem~\ref{thm:spectrum} gives $h_P(T)\le\sqrt{8(1+\pi^2)v_P M_P(T)}$.
Taking the maximum yields $h(T)\le\sqrt{8(1+\pi^2)v_{\mathrm{loc}}\mathcal M(T)}$.
Substitution into Eq.~\eqref{eq:mainbound} gives:
\begin{corollary}[Spectral-dephasing bound for whole-family interference]\label{cor:spectralhistory}
Under the state, projector, and waiting-time assumptions of Theorem~\ref{thm:whole}, for every $\delta>0$,
\begin{equation}\label{eq:directspectral}
\begin{aligned}
&\Prob\left(\max_{\alpha\ne\beta}|\epsilon_{\alpha\beta}|>\delta\right)\\
&\quad\le\frac{4\kappa(n-1)^2}{Q_*^2}\left(N_n+\frac{K_n}{\delta^2}\right)
\sqrt{8(1+\pi^2)v_{\mathrm{loc}}\mathcal M(T)}.
\end{aligned}
\end{equation}
In particular, for a sequence of such history families with fixed $n$, bounded $\kappa$, and uniformly positive $Q_*$, local Bohr-frequency dispersion $\mathcal M(T_D)\to0$ suffices for whole-family medium decoherence in probability.
\end{corollary}

Equation~\eqref{eq:directspectral} makes the local spectral-dephasing route to whole-family decoherence quantitative.
If every frequency window of width $1/T$ has small projector-weighted mass, the absolute time-averaged local correlation is small. This controls each conditional fluctuation and hence the normalized interference across the complete family with high probability.
This is a sufficient criterion for history decoherence. A family can also decohere through other mechanisms while its projectors retain memory.

\section{Spectral dephasing in translation-invariant lattices}\label{sec:static}

Translation-invariant lattices provide a many-body realization of local spectral dephasing. Translation averaging limits the zero-frequency spectral atom through the overlap of translated local supports. A bound on gap multiplicities controls the nonzero-frequency atoms. Together these estimates lead to weak long-time local memory along suitably chosen observation windows.

Let $d\ge1$ be the spatial dimension and $\Lambda=\prod_{j=1}^d\mathbb Z_{L_j}$ a periodic lattice of volume $V=|\Lambda|=\prod_{j=1}^d L_j$. Each site has a $q$-dimensional Hilbert space, so $D_V=q^V$. We write $h_{P,V}(T)$ for the memory parameter of a projector $P$ in this finite lattice. Let $U_{\mathbf x}$ implement a translation by $\mathbf x\in\Lambda$, and suppose $H_V$ commutes with every $U_{\mathbf x}$. The estimate uses translation symmetry, fixed local support, and factorization of the infinite-temperature trace on disjoint regions. The interaction range may be arbitrary.

Assume that the energy levels of $H_V$ are nondegenerate and that any nonzero energy difference $\omega$ occurs for at most $g_V$ ordered pairs of energy eigenstates. Let $P$ be a projector supported on a subset $S\subset\Lambda$, and define the number of translations for which its support overlaps itself by
\begin{equation}\label{eq:overlapcount}
C_S:=\bigl|\{\mathbf x\in\Lambda:S\cap(S+\mathbf x)\ne\varnothing\}\bigr|=|S-S|.
\end{equation}
For a fixed local shape, $C_S$ is bounded independently of volume. The overlap count controls the zero-frequency atom, and the gap-multiplicity assumption controls the nonzero-frequency atoms.

\begin{proposition}[Static-lattice long-time spectral-dephasing bound]\label{prop:static}
Under these assumptions, the zero-frequency and nonzero-frequency atoms of the local spectral measure satisfy
\begin{equation}\label{eq:atoms}
\mu_P(\{0\})\le\frac{C_Sv_P}{V},\qquad
\sup_{\omega\ne0}\mu_P(\{\omega\})\le\frac{g_V\|A_P\|_\infty^2}{D_V}.
\end{equation}
Moreover,
\begin{equation}\label{eq:staticbound}
\limsup_{T\to\infty}h_{P,V}(T)
\le 2\left[\frac{C_Sv_P}{V}+\sqrt{\frac{g_Vv_P}{D_V}}\right].
\end{equation}
\end{proposition}

\begin{proof}[Proof sketch]
Energy nondegeneracy restricts the zero-frequency atom to diagonal energy-basis matrix elements. Translation invariance lets us replace the centered projector in these elements by its spatial average. Translates with disjoint supports have zero infinite-temperature overlap, leaving at most $C_S$ contributing relative translations among the $V$ possibilities. At nonzero frequency, at most $g_V$ transitions contribute. Finally, the long-time mean square of the correlation is the sum of squared spectral-atom weights. Appendix~\ref{app:proof-static} gives the details.
\end{proof}

The geometric dependence is explicit. If $S$ lies inside a rectangular box of side lengths $\ell_1,\ldots,\ell_d$ with $L_j\ge2\ell_j$, then $C_S\le\prod_{j=1}^d(2\ell_j-1)$. In particular, for an $L^d$ cubic lattice and a box of side $\ell$, the zero-frequency bound is $\mu_P(\{0\})\le (2\ell-1)^dv_P/L^d$.

Consider a sequence of periodic lattices of fixed spatial dimension with $V\to\infty$ and $g_V/D_V\to0$. For a fixed finite collection of projectors with uniformly bounded support shapes, each $C_S=O(1)$ and $v_P\le1/4$. The long-time bound in Eq.~\eqref{eq:staticbound} then tends to zero. Choosing a sufficiently long window $T_V$ for each finite lattice makes all the corresponding $h_{P,V}(T_V)$ vanish along the sequence. Theorem~\ref{thm:spectrum} yields spectral nonconcentration at the same scales. For fixed history length, Theorem~\ref{thm:whole} then applies under uniform control of $\kappa$ and $Q_*^{-1}$.

The result establishes the existence of suitable observation windows. The atomic estimates alone give no uniform time scale for resolving nearly coincident frequencies. The related translation-invariant thermalization theorem of Pilatowsky-Cameo and Choi uses nonresonance ideas for different observables and initial states \cite{PilatowskyChoi2026}.

For a finite-size numerical illustration, specialize to the one-dimensional periodic spin-$1/2$ chain of $L$ sites, with $D_L=2^L$ and Hamiltonian

\begin{equation}\label{eq:model}
\begin{aligned}
H_L=\sum_{j=0}^{L-1}\bigl[&Z_jZ_{j+1}+X_j+\tfrac12 Z_j\\
&+\tfrac14(X_jY_{j+1}-Y_jX_{j+1})\bigr],
\end{aligned}
\end{equation}
Here $X_j$, $Y_j$, and $Z_j$ are Pauli operators on site $j$, and indices are understood modulo $L$.
The first term couples neighboring $Z$ components, the next two are transverse and longitudinal fields, and the final chiral term breaks reflection symmetry while preserving translations.
All coefficients are in the same energy units, with $\hbar=1$.
The spectra are numerically nondegenerate at the displayed sizes, although this does not establish nondegeneracy for arbitrary $L$.

At every observation we use the two-outcome projective decomposition $\mathcal P=\{P_0,P_1\}$, with $P_1=P=(\id+Z_0)/2$ and $P_0=\id-P$.
The initial state is maximally mixed, $\rho=\id/D_L$.
Thus the projector has normalized rank $q_P=1/2$, variance $v_P=1/4$, and centered part $A_P=Z_0/2$.
In particular, for a family of $n$ such binary observations, every reference history has weight $Q_\alpha=2^{-n}$, so $Q_*=2^{-n}$.
The transverse-field term ensures that $P$ is not conserved.
We use this same local decomposition in all spectral and history calculations, so the measured variable does not change between tests.

Exact diagonalization supplies the eigenvalues and matrix elements of $A_P$ needed for the weighted Bohr-frequency measure. The multitime-history calculations below instead use sparse Hamiltonian time evolution and computational-basis trace estimators.
For the zero-frequency check, we sum the diagonal energy-basis weights to obtain $\mu_P(\{0\})$.
For the nonzero-frequency diagnostic, we cluster computed nonzero Bohr gaps within an absolute frequency tolerance of $10^{-9}$ and sum their weights $|\langle m|A_P|l\rangle|^2/D_L$. Write $g_L^{\mathrm{num}}$ for the largest cluster multiplicity.
The computed values are $g_L^{\mathrm{num}}=1$ for $L=5,\ldots,8$, $g_L^{\mathrm{num}}=2$ for $L=9,10,11$, and $g_{12}^{\mathrm{num}}=3$. These tolerance-dependent clusters need not coincide with exact spectral atoms or exact energy-gap degeneracies.

\begin{figure*}[t]
\centering
\includegraphics[width=0.96\textwidth]{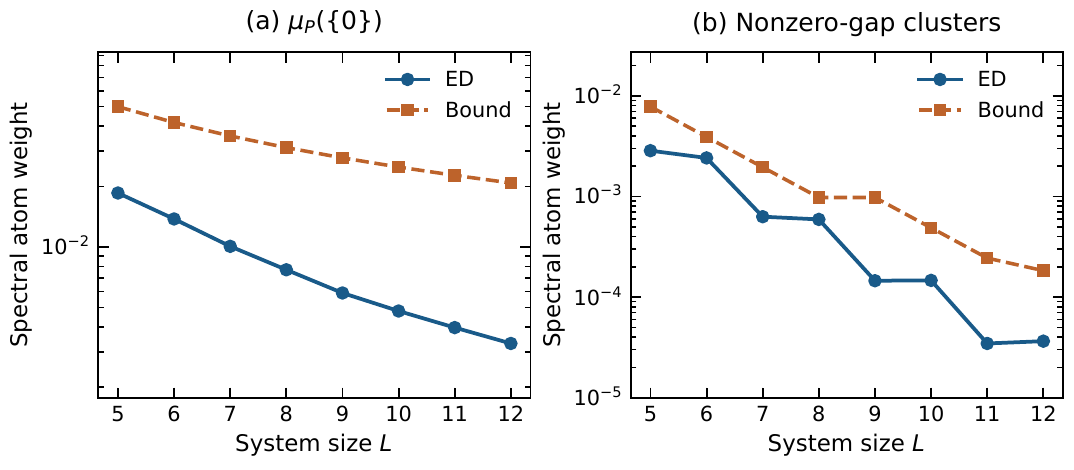}
\caption{Finite-size spectral diagnostics for the translation-invariant spin chain of Eq.~\eqref{eq:model}, $L=5,\ldots,12$. Solid circles denote exact-diagonalization (ED) data and dashed squares the corresponding bounds.
(a) Zero-frequency atom $\mu_P(\{0\})$ and spatial-averaging bound $v_P/L$, with $v_P=1/4$.
(b) Largest weight of a numerically grouped nonzero-frequency cluster and the cluster-multiplicity bound $g_L^{\mathrm{num}}\|A_P\|_\infty^2/D_L=g_L^{\mathrm{num}}/(4D_L)$. The absolute grouping tolerance is $10^{-9}$. The observed $g_L^{\mathrm{num}}$ is 1 for $L=5$--8, 2 for $L=9$--11, and 3 for $L=12$. Panel (b) is a finite-precision diagnostic of the nonzero-frequency estimate in Eq.~\eqref{eq:atoms}, not a determination of exact atom masses or gap degeneracies.}
\label{fig:zero}
\end{figure*}

The ED zero-frequency atoms satisfy the spatial-averaging bound, while the grouped nonzero-frequency weights satisfy the corresponding cluster-multiplicity estimates for the displayed sizes (Fig.~\ref{fig:zero}).
Proposition~\ref{prop:static} concerns exact spectral atoms. At finite numerical precision, grouping nearby frequencies provides a conservative diagnostic rather than a determination of the exact gap multiplicity.
The finite-size data do not determine the large-volume behavior of $g_L/D_L$ or the observation time required at larger sizes.

We now move beyond individual atoms to the finite-resolution spectral-dephasing criterion.
At $L=12$, we calculate $M_P(T)$ by sorting the weighted Bohr gaps from full double-precision exact diagonalization and scanning closed intervals of width $1/T$, without the gap-grouping tolerance used for Fig.~\ref{fig:zero}.
To estimate $h_P(T)$, we place these spectral weights on a frequency grid of spacing $4\times10^{-5}$, reconstruct the autocorrelation by an inverse fast Fourier transform, and apply a trapezoidal rule with 320 subintervals per observation window. The calculation uses 23 windows spanning $0.5\le T\le45$. Thus $M_P(T)$ uses the unbinned computed gaps, whereas $h_P(T)$ includes frequency-binning and quadrature approximations.

\begin{figure}[t]
\centering
\includegraphics[width=\columnwidth]{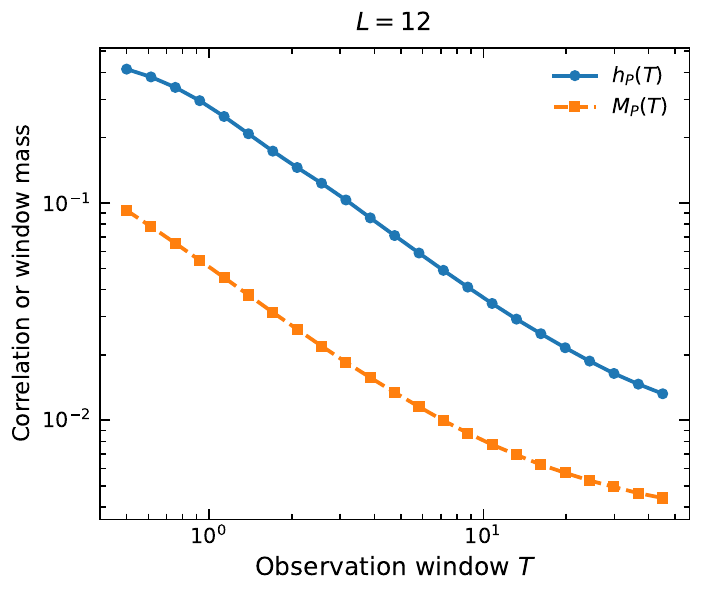}
\caption{Local Bohr-frequency window concentration $M_P(T)$ and time-averaged memory $h_P(T)$ for $L=12$ in the periodic spin-$1/2$ chain of Eq.~\eqref{eq:model}, with $P=(\id+Z_0)/2$ and $A_P=Z_0/2$. The $M_P(T)$ values use sliding windows on unbinned ED gaps. The memory parameter $h_P(T)$ is approximated using a frequency bin of $4\times10^{-5}$, an inverse FFT, and trapezoidal quadrature with 320 subintervals per window ($0.5\le T\le45$).}
\label{fig:spectral}
\end{figure}

Figure~\ref{fig:spectral} compares spectral concentration and dynamical memory at finite resolution.
The window mass $M_P(T)$ is nonincreasing as the resolution improves, while $h_P(T)$ may fluctuate. Theorem~\ref{thm:spectrum} relates their vanishing limits.
We next examine whether the same local measurement yields small interference across complete histories.

\begin{figure}[t]
\centering
\includegraphics[width=\columnwidth]{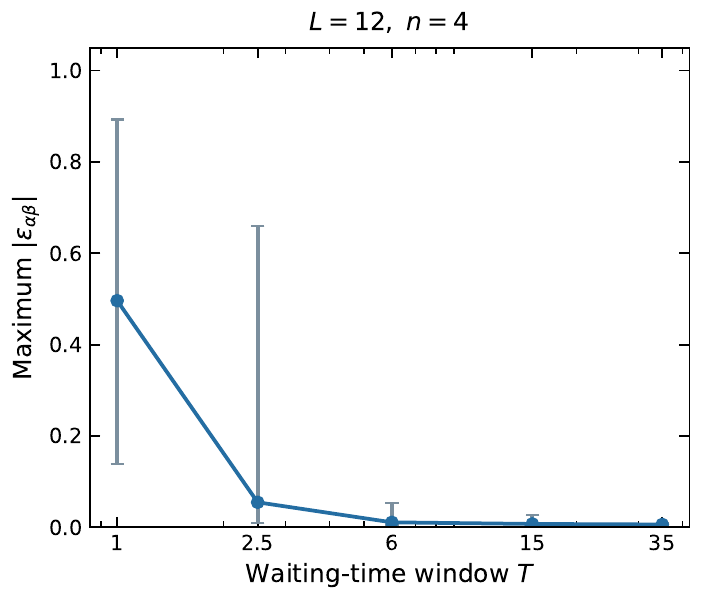}
\caption{Maximum estimated normalized interference across all 16 four-time ($n=4$) binary histories at $L=12$ in Eq.~\eqref{eq:model}. Each of the five windows $T=1,2.5,6,15,35$ contains 48 independent schedules with three waiting times sampled uniformly from $[0,T]$. For each schedule, 64 computational-basis probes are sampled without replacement per initial-outcome sector. Markers give the median, and vertical bars the tenth-to-ninetieth-percentile range of the estimates. This range includes probe-sampling variation and is not a confidence interval for the exact Gram matrix.}
\label{fig:hist}
\end{figure}

For the multitime numerical test, we take $L=12$ and $n=4$ observations (three independently sampled waiting times), giving $2^4=16$ binary histories.
We enumerate all 16 branches, but estimate their decoherence-functional overlaps without constructing dense chain operators. In each initial-outcome sector, computational-basis vectors are sampled uniformly without replacement and propagated through the sequence of projections using a sparse-matrix exponential.
The resulting unnormalized overlaps $\widehat D_{\alpha\beta}$ and branch weights $\widehat p_\alpha$ are trace estimators. Their normalized ratios $\widehat\epsilon_{\alpha\beta}=\widehat D_{\alpha\beta}/\sqrt{\widehat p_\alpha\widehat p_\beta}$ estimate the exact interference $\epsilon_{\alpha\beta}$. They are not exact Gram-matrix entries. For plotting, we use $\widehat G_{\alpha\beta}=\widehat D_{\alpha\beta}/\sqrt{\widehat p_\alpha\widehat p_\beta}$ when both estimated weights are positive, and zero otherwise. Its off-diagonal elements estimate $\epsilon_{\alpha\beta}$.
For Fig.~\ref{fig:hist}, each initial-outcome sector uses 64 probes. We report $\max_{\alpha\ne\beta}|\widehat\epsilon_{\alpha\beta}|$ for each of 48 independent time schedules per window, with independently sampled probes for every schedule. The resulting median and tenth and ninetieth percentiles therefore reflect both schedule variation and trace-sampling variation.

All positive estimated branch weights are retained in the normalization, although very small weights may be sensitive to finite arithmetic precision. The Hamiltonian is checked for Hermiticity and the estimated branch weights for normalization. For histories with different initial or final outcomes, the implementation imposes the exact zero overlap following from projector orthogonality.

Figure~\ref{fig:hist} shows the estimated multitime interference for the same projector used in the spectral and memory calculations.
As the allowed waiting-time window broadens from the short-time regime, the estimated maximum normalized interference typically decreases for the displayed $L=12$ chain.
Individual schedules may exhibit recurrences in a finite isolated system. The plotted spread also includes trace-estimator sampling variation.
Figure~\ref{fig:hist} illustrates the interference quantity controlled by Theorem~\ref{thm:whole}, rather than providing a direct numerical test of the quantitative probability bound in Eq.~\eqref{eq:mainbound}.

\begin{figure}[t]
\centering
\includegraphics[width=\columnwidth]{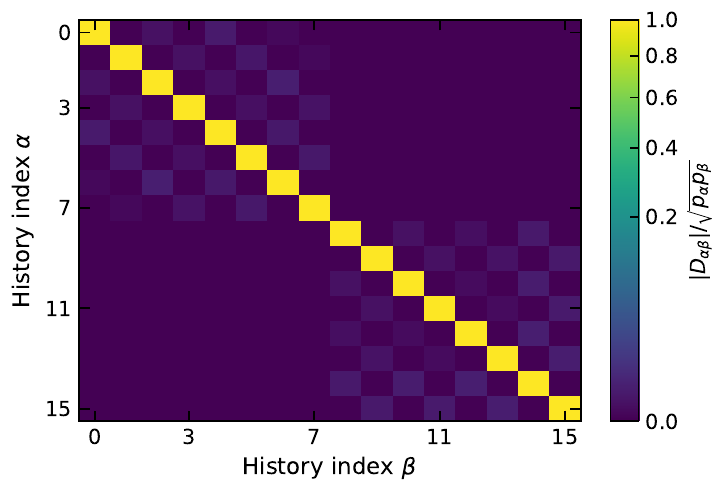}
\caption{Absolute values $|\widehat G_{\alpha\beta}|$ of the estimated normalized $16\times16$ history Gram matrix for $L=12$, $n=4$, and waiting times $(3.7,8.2,12.6)$ in the units of Eq.~\eqref{eq:model}. The trace estimator uses 256 uniformly sampled computational-basis probes without replacement per initial-outcome sector (out of 2048). Histories are ordered lexicographically, with outcomes $0$ and $1$ corresponding to $P_0=\id-P$ and $P_1=P$. Entries with different first or last outcomes are set to their exactly vanishing values by projector orthogonality. The largest estimated off-diagonal magnitude is $2.84\times10^{-3}$.}
\label{fig:gram}
\end{figure}

Figure~\ref{fig:gram} displays the modulus of the estimated normalized Gram matrix, rather than the complex Gram matrix itself. Its diagonal entries equal one by normalization for positive estimated branch weights.
For the maximally mixed initial state, different first or final outcomes imply exactly vanishing overlaps. Since those zero blocks are imposed by the calculation, they do not constitute an independent numerical test. The nontrivial entries are between histories sharing both endpoints but differing at intermediate observations.
For the displayed schedule the maximum estimated off-diagonal magnitude is about $2.84\times10^{-3}$. This is a finite-probe estimate without an accompanying probe-number convergence bound. Theorem~\ref{thm:whole} concerns the exact decoherence functional under its stated assumptions.

As a complementary analytic realization, Appendix~\ref{sec:dual} considers discrete-time dual-unitary Floquet circuits. Unlike the static-lattice setting, where local memory loss is established through spectral estimates in the long-time limit, these circuits provide an exact finite-time characterization of local memory. Before the light cone wraps around the system, exact two-point identities eliminate all nonzero-delay on-site correlations, yielding a discrete memory parameter $h_P^{\rm disc}(M)=v_P/M$. This allows the same local-memory-to-history argument to be applied without invoking the spectral assumptions of the static model, and gives an explicit $O(L^{-1})$ bound on the failure probability when $M\propto L$, under the stated conditions on the initial state and history weights. The Floquet example therefore demonstrates that suppression of interference across complete history families is not restricted to the static spectral setting, but can also follow from an independently established decay of local memory.

\section{Relation to other mechanisms of history decoherence}\label{sec:literature}
\subsection{Near-conservation versus spectral dephasing}
An established route to decoherent histories in isolated systems uses variables that remain conserved or approximately conserved on the observation time scale.
Brun and Halliwell obtained approximate decoherence for sufficiently coarse histories of a subsystem's total spin in a locally coupled chain \cite{BrunHalliwell1996}.
Halliwell linked histories of local densities to hydrodynamic equations, diffusion, and the approximate conservation of macroscopic number, momentum, and energy densities, as well as to a chain of coupled oscillators \cite{Halliwell1998,Halliwell1999Local,Halliwell2003}.
Here the slow evolution of the chosen variables preserves correlations useful for a quasiclassical, hydrodynamic picture.

Spectral dephasing instead concerns local observables with weak time-averaged connected correlations over the sampled waiting-time window.
A local observable with dispersed matrix-element-weighted Bohr frequencies can lose memory of its initial value. Under the assumptions of Theorem~\ref{thm:whole}, the branch weights approach the comparison products $Q_\alpha$, and the history Gram matrix becomes nearly diagonal.

Local memory loss is one sufficient mechanism for history decoherence.
Suppose a nontrivial projector $P$ commutes with a static Hamiltonian and the same decomposition is applied repeatedly.
Histories with changing outcomes have zero weight, and the remaining histories decohere exactly. Meanwhile, the spectral measure is $\mu_P=v_P\delta_0$, so $M_P(T)=v_P$ and $h_P(T)=2v_P$.
An exactly conserved projection can thus give decoherence through a mechanism different from dynamical memory loss \cite{HartleLaflammeMarolf1995}.

The number of observation times provides another basic distinction.
For $\rho=\id/D$ and only two observations, any unequal histories differ at an endpoint and therefore have exactly zero overlap.
At three or more observations, paths with the same endpoints can differ at intermediate times and interfere.
The nontrivial entries of Fig.~\ref{fig:gram} are examples of precisely this multitime structure.

\subsection{Multitime equilibration and higher-order correlations}
Multitime equilibration concerns process statistics, while history decoherence concerns interference between specified alternatives.
The finite-time theorems of Dowling and collaborators concern multitime process statistics with restrictions on initial states and on the temporal or spatial resolution of interventions \cite{Dowling2023Quantum,Dowling2023SciPost}.
Our theorem controls the normalized decoherence functional for a predetermined family of projective alternatives.
An equilibrated process may still retain quantum temporal memory, so the two conclusions must be established separately.

There are also mathematical approaches in which two-point clustering is used to infer higher-order connected-correlation properties.
Ampelogiannis and Doyon study such relations in $C^*$ dynamical systems \cite{AmpelogiannisDoyon2025}.
An entry of the history Gram matrix involves an alternating sequence of projectors and forward and backward propagators. This differs from an ordinary higher-order connected correlation.
The present argument succeeds because projectors are contractions, unequal outcomes create exact trace zeros, and each new waiting time is independent of the previous prefix.
These features let the frame estimate act on the complete prefix without assuming general higher-order factorization.

Consistency of history probabilities, classical Markovianity, and objective records impose different requirements \cite{Strasberg2023}.
The process-tensor framework tests memory against arbitrary sequences of operations \cite{Pollock2018,MilzModi2021}. Our result concerns the prescribed history family.
Influence matrices can possess extensive temporal entanglement even when simple local correlations have relaxed \cite{FolignoZhouBertini2023,Vilkoviskiy2026,WangVilkoviskiyAbanin2026}.
The fixed-history-length condition and the dependence on the number of branches therefore determine the reach of the estimate.
Random-matrix analyses of multipoint correlations provide yet another perspective on higher-order dynamical behavior \cite{Chorbadzhiyska2026}.

\subsection{Chaotic, integrable, and recorded histories}
Strasberg, Reinhard, and Schindler computed the decoherence functional directly for up to five-time histories in an isolated random-matrix model, finding strong suppression of interference for slow, coarse observables \cite{StrasbergReinhardSchindler2024}.
Their finite-size results support an exponential scaling conjecture, which differs in form from the sufficient probability bound proved here.
For Heisenberg chains, Wang and Strasberg compared chaotic, interacting integrable, and noninteracting integrable dynamics \cite{WangStrasberg2025}.
On nonequilibrium time scales they inferred strong exponential suppression for chaotic chains, weaker exponential suppression for interacting integrable chains, and no corresponding exponential suppression in the noninteracting case.
Their longer-time results reveal a different behavior for interacting integrable dynamics, with indications of slower suppression.
The theorem applies to each of these dynamical classes when the chosen local weighted Bohr spectra are sufficiently dispersed at the observation resolution. Establishing this property and its time scale in particular models remains a separate question.
The comparison suggests that the finite-time weighted Bohr spectrum may help distinguish the observed regimes without treating quantum chaos as an assumption.

Other models construct history alternatives through stored information.
Arefyeva and Polyakov identified sequential environmental modes that record the past of a locally coupled degree of freedom in a non-Markovian integrable environment, and demonstrated suppression of history interference numerically \cite{ArefyevaPolyakov2026}.
Fert\'e, Farci, and Cao found approximate decoherent histories of a coarse extensive observable in both the apparatus and scrambler phases of a solvable monitored system, with objective recording properties tied to the apparatus phase \cite{FerteFarciCao2026}.
These results sharpen the distinction between interference suppression and durable, locally accessible records, a distinction already important in the histories literature \cite{Halliwell1999}.
Our theorem concerns interference between histories of selected projectors and imposes no assumption about physical records.

The number of observations introduces another independent question.
In a random-matrix setting, Strasberg, Schindler, Wang, and Winter studied approximate records and observed structured recoherence for long histories \cite{StrasbergSchindlerWangWinter2026}.
Their results provide context for the explicit dependence on $n$ in Eq.~\eqref{eq:mainbound} and for the fixed-length scope of the present limiting statement.

\section{Discussion and conclusions}\label{sec:conclusion}
We have identified local spectral dephasing as a sufficient route to decoherent histories in isolated quantum many-body systems. Theorem~\ref{thm:spectrum} establishes the local criterion. Along sequences of systems and observation scales, projector-weighted spectral nonconcentration in windows of width $1/T$ is equivalent to vanishing absolute time-averaged connected autocorrelation. Both express the same asymptotic dephasing property. The equivalence concerns their vanishing limits and allows finite-time recurrences.

Theorem~\ref{thm:whole} and Corollary~\ref{cor:spectralhistory} connect this local property to the suppression of multitime interference. With independently sampled waiting times, a controlled initial-state operator norm, and positive reference branch weights, weak local memory suppresses every normalized off-diagonal overlap in a prescribed fixed-length history family with high probability. A uniform time-frame estimate applies after arbitrary earlier outcomes, and projective orthogonality controls overlaps between distinct branches. The same event yields approximate probability sum rules for every coarse graining of the family.

Translation-invariant static lattices supply explicit conditions for the local spectral premise. Proposition~\ref{prop:static} bounds the zero-frequency atom through translation averaging and the nonzero-frequency atoms through gap multiplicities. For fixed local support, nondegenerate energies, and $g_V/D_V\to0$, suitable volume-dependent observation windows yield vanishing spectral concentration and averaged local memory. Under the additional assumptions of the whole-family theorem, these systems therefore admit decoherent histories along the selected scales. The proposition establishes the existence of those windows without determining a universal finite-time rate.

For a one-dimensional spin chain, exact diagonalization provides the spectral comparisons, while sparse time evolution and computational-basis trace estimators illustrate interference across complete four-time history families. Appendix~\ref{sec:dual} gives a complementary discrete-time realization in dual-unitary Floquet circuits. Exact pre-wrap correlations give a local-memory parameter proportional to $1/M$, leading to an $O(L^{-1})$ failure-probability bound when $M\propto L$ and the state and branch-weight assumptions hold. Both examples enter the same local-memory-to-history argument.

Local spectral dephasing provides a route to history decoherence alongside hydrodynamic slow variables and persistent records. Its local two-time criterion does not require pointwise decay or Markovian dynamics under arbitrary interventions. Two questions remain particularly relevant: whether spectral nonconcentration can be proved on controlled, preferably polynomial, time scales for broad classes of local Hamiltonians, and whether the whole-family condition can be extended to histories whose length grows with system size while keeping small branch weights under control.

\appendix
\setcounter{corollary}{0}
\renewcommand{\thecorollary}{\Alph{section}\arabic{corollary}}
\renewcommand{\theHcorollary}{appendix.\Alph{section}.\arabic{corollary}}

\section{Proof of the finite-resolution spectral-dephasing criterion}\label{app:proof-spectral}
\begin{proof}[Proof of Theorem~\ref{thm:spectrum}]
For readability, write $\mu=\mu_P$, $v=v_P$, $M(T)=M_P(T)$, and $c(t)=c_P(t)$ within this proof.
The proof has two directions.
For the lower bound, we show that appreciable mass in one narrow frequency interval forces the absolute time average to remain appreciable.
For the upper bound, we show that if every such interval has little mass, then the squared correlation has a small time average, and hence so does its absolute value.

Introduce the nonnegative triangular function $f_T(t)=\max(1-|t|/T,0)$, supported on $[-T,T]$.
Its Fourier transform is the nonnegative squared sinc function:

\begin{equation}\label{eq:fejer}
\frac1T\int_{-T}^{T}f_T(t)\e^{\mathrm{i}xt}\,\mathrm{d}t=\operatorname{sinc}^2(xT/2),\qquad \operatorname{sinc}(y)=\frac{\sin y}{y}.
\end{equation}
Here $\operatorname{sinc}(0)$ is understood by continuity and equals one.
Take any frequency interval of width $T^{-1}$ with center $\Omega$.
Each frequency in that interval satisfies $|\omega-\Omega|T/2\le1/4$, so the squared sinc factor in Eq.~\eqref{eq:fejer} is at least $a_0$.
Now multiply $c(t)$ by $f_T(t)\e^{-\mathrm{i}\Omega t}$, integrate over $[-T,T]$, and divide by $T$.
Interchanging the time integral and the integral against $\mu$ is legitimate because $\mu$ is a finite positive measure.
The resulting frequency integral is nonnegative and at least $a_0$ times the mass of the chosen interval.
On the other hand, its absolute value cannot exceed $T^{-1}\int_{-T}^{T}|c(t)|\,\mathrm{d}t=h_P(T)$ because $0\le f_T\le1$.
Taking the supremum over centers gives $a_0M(T)\le h_P(T)$.

For the upper bound, first apply Cauchy--Schwarz on the interval of length $2T$ to obtain $h_P(T)^2\le(2/T)\int_{-T}^{T}|c(t)|^2\,\mathrm{d}t$.
The triangular weight $f_{2T}$ is at least $1/2$ throughout $[-T,T]$.
Using this weight and expanding $|c(t)|^2$ as a double integral over the positive measure $\mu$, the same Fourier identity at scale $2T$ gives

\begin{equation}\label{eq:squarefejer}
h_P(T)^2\le8\int\!\!\int \operatorname{sinc}^2\!\bigl(T(\omega-\omega')\bigr)\,\mu(\mathrm{d}\omega)\mu(\mathrm{d}\omega').
\end{equation}
It remains to control the double integral using only the maximum window mass.
Fix $\omega$ and partition the $\omega'$-axis into half-open intervals $I_j=[\omega+(j-1/2)/T,\omega+(j+1/2)/T)$ for integers $j$.
Each has width $1/T$, so its $\mu$-mass is at most $M(T)$, irrespective of where it sits in the spectrum.
For $j=0$, the squared sinc factor is at most one. For $j\ne0$, its argument has magnitude at least $|j|-1/2$, so the factor is bounded by $(|j|-1/2)^{-2}$.
Summing the contributions of these intervals yields

\begin{equation}\label{eq:sumfejer}
\begin{aligned}
&\int \operatorname{sinc}^2\!\bigl(T(\omega-\omega')\bigr)\,\mu(\mathrm{d}\omega')\\
&\quad\le\left[1+2\sum_{j=1}^{\infty}(j-1/2)^{-2}\right]M(T)\\
&\quad=(1+\pi^2)M(T).
\end{aligned}
\end{equation}
The numerical constant follows from the convergent series $\sum_{j=1}^{\infty}(j-1/2)^{-2}=\pi^2/2$.
The remaining integral over $\omega$ has total mass $v$, so Eqs.~\eqref{eq:squarefejer} and \eqref{eq:sumfejer} give $h_P(T)^2\le8(1+\pi^2)vM(T)$, which is the desired upper bound.
The proof allows coincident frequencies, including the zero-frequency atom.\qedhere

\end{proof}

\section{Proof of the complete-history interference bound}\label{app:proof-whole}
\begin{proof}[Proof of Theorem~\ref{thm:whole}]
We first prove a one-step inequality that is independent of the history construction.
Let $A$ be any centered Hermitian operator, with $\Tr A=0$, and let $A(t)=U(t)^\dagger A U(t)$.
Define $c_A(t)=D^{-1}\Tr[A(t)A]$ and $h_A(T)=T^{-1}\int_{-T}^{T}|c_A(t)|\,\mathrm{d}t$, in analogy with Eq.~\eqref{eq:mixing}.
For any matrix $X$, let $\|X\|_2^2=\Tr(X^\dagger X)$ denote its squared Hilbert--Schmidt norm.
Without a Hermiticity or locality assumption on $X$, the time-frame estimate is

\begin{equation}\label{eq:frame}
\frac1T\int_0^T|\Tr[XA(t)]|^2\,\mathrm{d}t\le D\|X\|_2^2h_A(T),
\end{equation}
To see the estimate, equip matrix space with the inner product $\langle B,C\rangle=\Tr(B^\dagger C)$ and let $\mathcal F$ map a matrix $Y$ to the function $(\mathcal FY)(t)=\Tr[A(t)Y]$ on $[0,T]$.
The codomain has the normalized inner product associated with $L^2([0,T],\mathrm{d}t/T)$, so $\|\mathcal FX\|^2$ is exactly the left side of Eq.~\eqref{eq:frame}.
For a function $f$ in the codomain, the adjoint is $\mathcal F^\dagger f=T^{-1}\int_0^T f(t)A(t)\,\mathrm{d}t$.
Consequently, $\mathcal F\mathcal F^\dagger$ acts by an integral kernel which, relative to ordinary Lebesgue measure, is

\begin{equation}\label{eq:kernel}
K(s,t)=\frac1T\Tr[A(s)A(t)]=\frac{D}{T}c_A(s-t).
\end{equation}
For fixed $s\in[0,T]$, the integration variable $t\in[0,T]$ makes $s-t$ range within $[-T,T]$.
The absolute row integral of this kernel is therefore no larger than $D h_A(T)$, and stationarity gives the same bound for every column.
Schur's test bounds the operator norm $\|\mathcal F\mathcal F^\dagger\|$ by $D h_A(T)$.
Since $\mathcal F^\dagger\mathcal F$ and $\mathcal F\mathcal F^\dagger$ have the same nonzero spectrum, the claimed inequality follows.

Two features of Eq.~\eqref{eq:frame} are essential for what follows.
First, it uses only the two-time correlation of $A$, through the translation-invariant kernel $c_A(s-t)$.
Second, it holds uniformly for arbitrary $X$, even if $X$ is a complicated non-Hermitian operator encoding all earlier branches.
The same proof applies with $A(-t)$ in place of $A(t)$, because the autocorrelation is even.

We now propagate this one-step estimate through the history.
For an ordered pair $\alpha,\beta$, let $X_k^{\alpha\beta}$ be the operator obtained by evolving both prefixes through observation $k$ (with the two outcome projectors acting on its left and right).
At the initial observation, set $X_0^{\alpha\beta}=P^{(0)}_{a_0}\rho P^{(0)}_{b_0}$. For $k\ge1$, define

\begin{equation}\label{eq:xrec}
X_k^{\alpha\beta}=P_{a_k}^{(k)}U(\tau_k)X_{k-1}^{\alpha\beta}U(\tau_k)^\dagger P_{b_k}^{(k)}.
\end{equation}
Write $d_k^{\alpha\beta}=\Tr X_k^{\alpha\beta}$ for the trace of this prefix.
At the final observation, $d_{n-1}^{\alpha\beta}=D_{\alpha\beta}$, with $p_\alpha=d_{n-1}^{\alpha\alpha}$ on the diagonal.
Every unitary preserves the Hilbert--Schmidt norm, and multiplying by a projector on either side cannot increase it.
Thus $\|X_k^{\alpha\beta}\|_2\le\|\rho\|_2$ for all pairs and all $k$.
Moreover, $\Tr\rho^2\le\|\rho\|_\infty\Tr\rho\le\kappa/D$, so $\|X_k^{\alpha\beta}\|_2^2\le\kappa/D$.
This is where the initial-state hypothesis enters.

Next consider the trace at observation $k\ge1$.
If the two histories choose different outcomes, $a_k\ne b_k$, orthogonality $P_{b_k}^{(k)}P_{a_k}^{(k)}=0$ and cyclicity of the trace give $d_k^{\alpha\beta}=0$ exactly, irrespective of what happened earlier.
If they choose the same outcome $a_k=b_k=a$, cyclicity removes one of the two equal projectors from the trace.
Decomposing the remaining projector as $P_a^{(k)}=q_a^{(k)}\id+A_{P_a^{(k)}}$ then yields the exact recursion

\begin{equation}\label{eq:recursion}
\begin{aligned}
d_k^{\alpha\beta}&=q_a^{(k)}d_{k-1}^{\alpha\beta}+e_k^{\alpha\beta},\\
e_k^{\alpha\beta}&=\Tr[X_{k-1}^{\alpha\beta}A_{P_a^{(k)}}(\tau_k)].
\end{aligned}
\end{equation}

The first term propagates the previous trace by the normalized rank $q_a^{(k)}$, while $e_k^{\alpha\beta}$ is the fluctuation due to the dynamically evolved centered projector.
Crucially, after conditioning on $\tau_1,\ldots,\tau_{k-1}$, the operator $X_{k-1}^{\alpha\beta}$ is fixed, whereas $\tau_k$ is still uniform on $[0,T]$ and independent of that prefix.
Applying Eq.~\eqref{eq:frame} with $X=X_{k-1}^{\alpha\beta}$ and $A=A_{P_a^{(k)}}$, then using $D\|X\|_2^2\le\kappa$ and the definition of $h(T)$, gives

\begin{equation}\label{eq:evariance}
\Ex\bigl[|e_k^{\alpha\beta}|^2\,\big|\,\tau_1,\ldots,\tau_{k-1}\bigr]\le\kappa h(T).
\end{equation}

We next turn the one-step estimate into bounds on final traces.
For a diagonal pair, $d_0^{\alpha\alpha}=r_{a_0}$.
If all fluctuations $e_k^{\alpha\alpha}$ were absent, repeated multiplication by the $q_{a_k}^{(k)}$ would give exactly $Q_\alpha$.
The difference $p_\alpha-Q_\alpha$ is therefore a sum of at most $n-1$ fluctuations, each multiplied by a product of later normalized ranks, all of which lie in $[0,1]$.

For an off-diagonal pair $\alpha\ne\beta$, let $j$ be the last observation at which the outcomes differ, including $j=0$ if only the initial outcomes differ.
At that observation $d_j^{\alpha\beta}=0$: for $j\ge1$ this follows from projector orthogonality in the recursion, and for $j=0$ it follows directly from $\Tr(P_{a_0}^{(0)}\rho P_{b_0}^{(0)})=0$.
All observations after $j$ have matching outcomes, so only fluctuations from the remaining suffix can contribute to the final overlap.
Their coefficients again have magnitude at most one.
Thus both the deviation of a diagonal weight from $Q_\alpha$ and any unequal-history overlap are bounded by sums of at most $n-1$ terms.
The elementary inequality $|\sum_{k=1}^m z_k|^2\le m\sum_{k=1}^m|z_k|^2$, together with Eq.~\eqref{eq:evariance} and the law of total expectation, yields

\begin{equation}\label{eq:secondmoments}
\begin{aligned}
\Ex|p_\alpha-Q_\alpha|^2&\le\kappa(n-1)^2h(T),\\
\Ex|D_{\alpha\beta}|^2&\le\kappa(n-1)^2h(T)\quad(\alpha\ne\beta).
\end{aligned}
\end{equation}
Notice that the different fluctuations need not be mutually independent.
Independence of each new waiting time from the preceding prefix is sufficient for the conditional estimate. The final step uses a deterministic inequality for a sum.

Finally, apply Markov's inequality to the nonnegative squared deviations (equivalently, Chebyshev's inequality in this form).
For any one history, the probability that $|p_\alpha-Q_\alpha|>Q_*/2$ is at most $4\kappa(n-1)^2h(T)/Q_*^2$.
Summing over $N_n$ branches bounds the probability that any branch violates this inequality.
Similarly, each off-diagonal condition $|D_{\alpha\beta}|>\delta Q_*/2$ has probability at most $4\kappa(n-1)^2h(T)/(\delta^2Q_*^2)$, and there are $K_n$ ordered unequal pairs.
Outside the union of these two exceptional events, every branch obeys $p_\alpha\ge Q_*/2$, so $\sqrt{p_\alpha p_\beta}\ge Q_*/2$ and $|\epsilon_{\alpha\beta}|\le\delta$ for all unequal pairs.
Adding the two union bounds gives Eq.~\eqref{eq:mainbound} and the stated simultaneous conclusion.

\end{proof}

\section{Proof of the static-lattice spectral bound}\label{app:proof-static}
\begin{proof}[Proof of Proposition~\ref{prop:static}]
Let $\Lambda=\prod_{j=1}^d\mathbb Z_{L_j}$, $V=|\Lambda|$, and $D_V=q^V$ be as in Sec.~\ref{sec:static}. Put $A=P-q_P\id$, write $A_{ml}=\langle m|A|l\rangle$, and abbreviate $\mu=\mu_P$ and $c=c_P$. Let $U_{\mathbf x}$ denote translation by $\mathbf x\in\Lambda$. We keep the spatial dimension $d\ge1$ general throughout the proof.

We first bound the zero Bohr frequency. Because the energy spectrum is nondegenerate, $E_m-E_l=0$ implies $m=l$. Since $H_V$ commutes with each $U_{\mathbf x}$, every energy eigenstate is also a translation eigenstate. Define the translation average
\begin{equation}\label{eq:spatialaverage}
\overline A=\frac1V\sum_{\mathbf x\in\Lambda}U_{\mathbf x}AU_{\mathbf x}^\dagger.
\end{equation}
Translation phases cancel in diagonal energy-basis matrix elements, so $\langle m|A|m\rangle=\langle m|\overline A|m\rangle$. Therefore,
\begin{equation}\label{eq:zeroaverage}
\mu(\{0\})=\frac1{D_V}\sum_m|A_{mm}|^2
\le\frac1{D_V}\Tr\overline A^2.
\end{equation}

Expanding the square and using cyclicity of the trace and translation invariance gives
\begin{equation}\label{eq:transavg}
\frac1{D_V}\Tr\overline A^2
=\frac1V\sum_{\mathbf x\in\Lambda}
\frac1{D_V}\Tr\bigl(AU_{\mathbf x}AU_{\mathbf x}^\dagger\bigr)
\le\frac{C_S}{V}v_P.
\end{equation}
Indeed, $A$ is traceless and supported on $S$. If $S$ and $S+\mathbf x$ are disjoint, the normalized trace in the sum factorizes and vanishes. There are exactly $C_S=|S-S|$ relative translations for which the supports intersect. For each of them, normalized Hilbert--Schmidt Cauchy--Schwarz yields
\begin{equation}\label{eq:translatecs}
\left|\frac1{D_V}\Tr\bigl(AU_{\mathbf x}AU_{\mathbf x}^\dagger\bigr)\right|
\le\frac1{D_V}\Tr A^2=v_P.
\end{equation}
Equations~\eqref{eq:zeroaverage} and \eqref{eq:transavg} prove the zero-frequency part of Eq.~\eqref{eq:atoms}. If $S$ is contained in a box with side lengths $\ell_j$ and $L_j\ge2\ell_j$, then all overlapping translation vectors lie in the corresponding coordinatewise difference box, giving $C_S\le\prod_{j=1}^d(2\ell_j-1)$. In one dimension this reduces to $2\ell-1$, and for a cubic system it is bounded by $(2\ell-1)^d$.

For any nonzero Bohr frequency, the gap assumption bounds the number of contributing ordered energy pairs by $g_V$. Since $|A_{ml}|\le\|A\|_\infty$,
\begin{equation}\label{eq:gapatom}
\mu(\{\omega\})=\frac1{D_V}\sum_{(m,l):E_m-E_l=\omega}|A_{ml}|^2
\le\frac{g_V}{D_V}\|A\|_\infty^2,\qquad\omega\ne0.
\end{equation}
This is the other part of Eq.~\eqref{eq:atoms} and does not depend on lattice dimension.

To derive the long-time memory bound, fix the finite lattice $\Lambda$ first. Its Bohr-frequency measure has finitely many atoms, and the oscillatory cross terms between distinct atoms average to zero. Thus the finite-spectrum Wiener identity reads
\begin{equation}\label{eq:wiener}
\lim_{T\to\infty}\frac1T\int_{-T}^{T}|c(t)|^2\,\mathrm{d}t
=2\sum_{\omega}\mu(\{\omega\})^2.
\end{equation}
By Cauchy--Schwarz on $[-T,T]$ and the normalization of $h_{P,V}(T)$, we obtain
\begin{equation}\label{eq:wienermemory}
\limsup_{T\to\infty}h_{P,V}(T)
\le2\sqrt{\sum_{\omega}\mu(\{\omega\})^2}.
\end{equation}
Splitting off zero frequency and using $\sum_{\omega}\mu(\{\omega\})=v_P$ gives
\begin{equation}\label{eq:atomssquared}
\sum_{\omega}\mu(\{\omega\})^2
\le \left(\frac{C_Sv_P}{V}\right)^2
+\frac{g_V\|A\|_\infty^2}{D_V}\,v_P.
\end{equation}
Finally $\sqrt{x^2+y^2}\le x+y$ for nonnegative $x,y$, and $\|A\|_\infty\le1$ for a centered projector. Combining Eqs.~\eqref{eq:wienermemory} and \eqref{eq:atomssquared} yields Eq.~\eqref{eq:staticbound}.

The limits are taken in this order: first $T\to\infty$ at each fixed finite volume, then $V\to\infty$ under the stated support and gap-multiplicity assumptions. 
\end{proof}

\section{Discrete-time realization: dual-unitary Floquet circuits}\label{sec:dual}
This appendix considers a complementary discrete-time driven system: a dual-unitary brickwork circuit \cite{BertiniKosProsen2019,ClaeysLamacraft2021,PiroliBertiniCiracProsen2020}.
Here one Floquet period consists of two staggered layers of a fixed two-site gate on an even periodic chain.
The gate is unitary both in the ordinary time-evolution pairing of indices and in the pairing obtained by interchanging the spatial and temporal directions.
Let $F$ denote the unitary evolution through one complete period.
For the centered part $A_P=P-q_P\id$ of an on-site projector $P$, write $A_P(m)=F^{-m}A_PF^m$ and $c_P(m)=D^{-1}\Tr[A_P(m)A_P]$ at integer period $m$.
This notation is the discrete counterpart of Eq.~\eqref{eq:mixing}, not an assertion that the circuit has a static Hamiltonian of the type used in Sec.~\ref{sec:spectral}.

In dual-unitary dynamics, exact light-cone identities force the infinite-temperature on-site autocorrelation of a traceless one-site operator to vanish at nonzero integer delays. This holds before the causal cone wraps around the periodic chain.
Here $M$ is the largest sampled number of Floquet periods. The symbol is distinct from the spectral concentration $M_P(T)$ defined in Sec.~\ref{sec:spectral}.
For definiteness, we impose the conservative condition $4M<L$ on this window: the two edges of the depth-$2M$ causal cone then cannot meet around the ring.
As a discrete version of the absolute-memory parameter, define

\begin{equation}\label{eq:discmix}
h_P^{\rm disc}(M)=\frac1M\sum_{r=-(M-1)}^{M-1}|c_P(r)|.
\end{equation}
The sum covers every possible difference of two sampled waiting times in $\{1,\ldots,M\}$.
It includes $r=0$, which is essential: this diagonal Gram-matrix term remains equal to $c_P(0)=v_P$ even when all nonzero-delay correlations vanish.
Thus, for independent waiting times uniform on $\{1,\ldots,M\}$ and the pre-wrap on-site identity $c_P(r)=0$ for $0<|r|<M$, one has exactly $h_P^{\rm disc}(M)=v_P/M$.
The next result substitutes this discrete memory scale for the continuous $h(T)$ in the whole-family theorem.
Define the maximal variance of the positive-time projectors as $v_*=\max_{1\le k\le n-1,\,a\in\mathcal A_k}v_{P_a^{(k)}}=\max_{k,a}q_a^{(k)}(1-q_a^{(k)})$.

\begin{corollary}[Pre-wrap Floquet histories]\label{cor:dual}
Retain $n\ge2$, the predetermined complete projective decompositions, the state bound $\rho\le\kappa\id/D$, and $Q_*>0$ from Theorem~\ref{thm:whole}.
Suppose that all projectors after the initial time are one-site projectors in a dual-unitary circuit satisfying the on-site two-point identity above, with an even ring length $L$ and $4M<L$.
For independently uniform integer waiting times between one and $M$ Floquet periods, Eq.~\eqref{eq:mainbound} holds with $h(T)$ replaced by $v_*/M$.
For fixed history length and size-independent normalized one-site projector ranks, uniformly bounded $\kappa$, and a size-independent lower bound $Q_*\ge q_*>0$, choosing $M=\lfloor cL\rfloor$ with fixed $0<c<1/4$ gives, for sufficiently large $L$, an $O(L^{-1})$ upper bound on the failure probability.
\end{corollary}

\begin{proof}
The argument repeats the continuous time-frame step with integrals replaced by finite sums.
For a centered Hermitian operator $A$, set $A(m)=F^{-m}AF^m$ and $c_A(r)=D^{-1}\Tr[A(r)A]$, and define the linear map $\mathcal F_M$ whose $m$th component is $\Tr[A(m)X]/\sqrt M$, for $m=1,\ldots,M$.
The Gram matrix of this map has entry $D c_A(s-m)/M$ in row $s$ and column $m$.
Since each row and each column sums over a subset of the differences $-(M-1),\ldots,M-1$, the discrete Schur bound yields

\begin{equation}\label{eq:discframe}
\frac1M\sum_{m=1}^{M}|\Tr[X A(m)]|^2\le D\|X\|_2^2\frac1M\sum_{r=-(M-1)}^{M-1}|c_A(r)|.
\end{equation}
The term $r=0$ necessarily contributes to the Gram-matrix diagonal.
For the traceless centered operators $A_P=P-q_P\id$ of one-site projectors, the pre-wrap identity eliminates all nonzero differences, leaving $h_P^{\rm disc}(M)=v_P/M$.
At each history step, independence of the new integer waiting time permits the same conditional second-moment estimate as before, now using Eq.~\eqref{eq:discframe}.
The trace recursion, its contraction factors, and the final union bound are unchanged.
Substituting $v_*/M$ for $h(T)$ proves the corollary.

\end{proof}

This example makes the source of the suppression especially transparent.
In the pre-wrap window, nonzero-lag on-site two-point correlations vanish exactly, and the residual discrete frame parameter is solely the equal-time contribution $v_P/M$.
Choosing $M=\lfloor cL\rfloor$ with $0<c<1/4$ therefore produces a failure-probability bound proportional to $1/L$, provided the other factors in the theorem are controlled.
In particular, fixed normalized ranks for on-site projectors and the upper bound on $\rho$ alone do not guarantee that the initial outcome probabilities $r_{a_0}$, and hence $Q_*$, stay bounded away from zero.
The explicit $Q_*$ assumption in the corollary cannot be omitted from that rate claim.

The circuit result uses an exact discrete-time two-point identity. It enters the conditional-history recursion in the same way as the spectral memory estimate for static lattices.
The two-point identity alone leaves higher-order temporal correlations and arbitrary operation sequences unrestricted.

\bibliographystyle{apsrev4-2}
\bibliography{references}
\end{document}